\documentclass[aps, prb, superscriptaddress, twocolumn]{revtex4}
\usepackage{amssymb}
\usepackage{graphicx}
\usepackage{hyperref}

\begin{document}

\title{Crystallography, magnetic susceptibility, heat capacity, and electrical resistivity of heavy fermion LiV$_2$O$_4$ single crystals grown using a self-flux technique}
\author{S. Das}
\affiliation{Ames Laboratory and Department of Physics and Astronomy, Iowa State University, Ames, Iowa 50011}
\author{X. Zong}
\affiliation{Ames Laboratory and Department of Physics and Astronomy, Iowa State University, Ames, Iowa 50011}
\author{A. Niazi}
\affiliation{Ames Laboratory and Department of Physics and Astronomy, Iowa State University, Ames, Iowa 50011}
\author{A. Ellern}
\affiliation{Department of Chemistry, Iowa State University, Ames, Iowa 50011}
\author{J. Q. Yan}
\affiliation{Ames Laboratory and Department of Physics and Astronomy, Iowa State University, Ames, Iowa 50011}
\author{D. C. Johnston}
\affiliation{Ames Laboratory and Department of Physics and Astronomy, Iowa State University, Ames, Iowa 50011}

\date{\today}
    
\begin{abstract}

Magnetically pure spinel compound ${\rm LiV_2O_4}$ is a rare $d$-electron heavy fermion. Measurements on single crystals are needed to clarify the mechanism for the heavy fermion behavior in the pure material.  In addition, it is known that small concentrations ($< 1$ mol\%) of magnetic defects in the structure strongly affect the properties, and measurements on single crystals containing magnetic defects would help to understand the latter behaviors.  Herein, we report flux growth of ${\rm LiV_2O_4}$ and preliminary measurements to help resolve these questions. The magnetic susceptibility of some as-grown crystals show a Curie-like upturn at low temperatures, showing the presence of magnetic defects within the spinel structure. The magnetic defects could be removed in some of the crystals by annealing them at 700 $^\circ$C\@. A very high specific heat coefficient $\gamma$ = 450 mJ/(mol K${^2}$\@) was obtained at a temperature of 1.8 K for a crystal containing a magnetic defect concentration $n$${\rm _{defect}}$ = 0.5 mol\%. A crystal with $n$${\rm _{defect}}$ = 0.01 mol\% showed a residual resistivity ratio of 50.

\end{abstract}

\maketitle
\section{\label{intro}introduction}

The spinel lithium vanadium oxide ${\rm LiV_2O_4}$ is a material of great interest as it shows heavy fermion behavior\cite{Kondo1997, Johnston2000} in spite of being a $d$-electron metal whereas most of the other heavy fermions are $f$-electron compounds. The origin of this heavy fermion behavior in ${\rm LiV_2O_4}$ is controversial. ${\rm LiV_2O_4}$ has the normal face-centered-cubic spinel structure with the space group \textit{Fd$\overline{3}$m}. The V atoms are coordinated by six O atoms in a slightly distorted octahedron. The Li atoms are coordinated with four O atoms in a tetrahedron. The Li atoms are located in the gaps between chains of the ${\rm VO_6}$ edge-sharing octahedra.  A study of the phase relations in the Li$_2$0-V$_2$O$_3$-V$_2$O$_5$ system at 700~$^\circ$C  (Ref.~\onlinecite{Das}) showed that the homogeneity range of ${\rm LiV_2O_4}$ is smaller than the experimental resolution of $\sim 1$~at\%.  From NMR measurements done on ${\rm LiV_2O_4}$ samples it has been found that for magnetically pure samples the $^{7}$Li nuclear spin-lattice relaxation rate ${1/T_1}$ is proportional to temperature \textit{T} at low temperatures (the Korringa law) which is typical for Fermi liquids.\cite{Kondo1997, Mahajan1998, Fujiwara2004} However for samples which contain magnetic defects within the spinel structure, the relaxation rate has a peak at $\sim1$ K and also shows other signatures which do not agree with the behavior of Fermi liquids.\cite{Johnston2005,Kaps} The occurrence of magnetic defects is easily seen as a low-$T$ Curie-like upturn in the magnetic susceptibility rather than becoming nearly independent of $T$ below $\sim$ 10 K as observed for the intrinsic behavior.\cite{Kondo1999} We have proposed a model in which the magnetic defects arise from a small homogeneity range of ${\rm LiV_2O_4}$ in the spinel structure.\cite{Das} High quality crystals containing magnetic defects might help to resolve the question of the nature of these defects and may shed light on the mechanism for heavy fermion behavior in the pure material and on whether a Fermi liquid is still present in samples containing magnetic defects. In particular, there may be a critical concentration separating Fermi liquid from non-Fermi liquid behaviors.

Crystal growth reports of ${\rm LiV_2O_4}$ are rare. ${\rm LiV_2O_4}$ crystals were first grown by hydrothermal reaction of ${\rm LiVO_2}$ and ${\rm VO_2}$ in aqueous solutions 1N in ${\rm LiOH}$ sealed in gold tubes and heated to 500 -- 700~$^\circ$C under a pressure of 3~kbar for $\sim$ 24 hr.\cite{Rogers} Octahedra shaped  crystals were obtained that were $\sim$ 0.75~mm on an edge. Electrical resistivity measurements demonstrated for the first time that ${\rm LiV_2O_4}$ is metallic down to a temperature $T$ of at least 4~K, with a room temperature resistivity of 300 to 800~$\mu$$\Omega$ cm depending on the crystal.\cite{Rogers} Electrical resistivity measurements of magnetically pure ${\rm LiV_2O_4}$ single crystals using crystals grown by this technique were recently reported\cite{Takagi,Urano} down to 0.3~K revealing a $T^2$ dependence between 0.3 and $\sim 2$~K as expected for a Fermi liquid.  Heat capacity ($C$) measurements on these crystals yielded an extrapolated zero-temperature $C/T$ value of 350~mJ/mol~K$^2$ which was comparable to the value of $C/T$ $\sim$ 430~mJ/mol~K$^2$ previously obtained at 1~K from measurements on polycrystalline samples.\cite{Johnston1999, Kondo1997} More recently, the first flux growth of single crystals of ${\rm LiV_2O_4}$ was reported using ${\rm LiCl-Li_2MoO_4-LiBO_2}$ as the flux.\cite{Matsushita} The crystals were reported to be of high quality but extremely reactive to air and/or moisture.\cite{Matsushita}

In this paper we report a new self-flux growth method to obtain single crystals of ${\rm LiV_2O_4}$ along with our initial magnetic, thermal, and transport properties of our crystals. Some of our as-grown crystals had magnetic defects in them while some were essentially defect free. Unlike the crystals grown in Ref.~[\onlinecite{Matsushita}], our crystals did not show any reactivity towards air and moisture.

\section{\label{expt}Experimental Details}

The starting materials of our samples of ${\rm LiV_2O_4}$ and ${\rm Li_3VO_4}$ were ${\rm Li_2CO_3}$ (99.995\%, Alfa Aesar), ${\rm V_2O_5}$ (99.999\%, M V Laboratories Inc.), and ${\rm V_2O_3}$ (99.999\%, M V Laboratories Inc.). The crystals of ${\rm LiV_2O_4}$ were grown in a vertical tube furnace. The single crystal X-ray diffraction measurements were done using a Bruker CCD-1000 diffractometer with Mo K$_{\alpha}$ ($\lambda$ = 0.71073 \AA) radiation. Powder X-ray diffraction measurements at room temperature were done using a Rikagu Geigerflex diffractometer with a curved graphite crystal monochromator. Differential thermal analysis experiments were carried out using a Perkin-Elmer differential thermal analyzer (DTA-6). The magnetic measurements on the crystals were done using a Quantum Design superconducting quantum interference device (SQUID) magnetometer in the temperature range 1.8 K -- 350 K and magnetic field range 0 -- 5.5 T\@. The heat capacity and electrical resistivity measurements were done using a Quantum Design physical property measurement system (PPMS). For the heat capacity measurements, Apiezon N grease was used for thermal coupling between the samples and the sample platform. Heat capacity was measured in the temperature range 1.8 K -- 300~K\@. For electrical resistivity measurements, 0.001~inch diameter platinum (99.999\%) leads were put on polished crystals using single component Epotek P1011 epoxy glue for electrical connections. Electrical resistivity was measured in the temperature range 1.8 K -- 300~K in 0 and 5~T magnetic fields.

\section{\label{growth}Crystal Growth and Characterization}

\subsection{\label{expt}LiV$_2$O$_4$ -- Li$_3$VO$_4$ Pseudobinary Phase Diagram}

\begin{figure}
\includegraphics[width=2.5in]{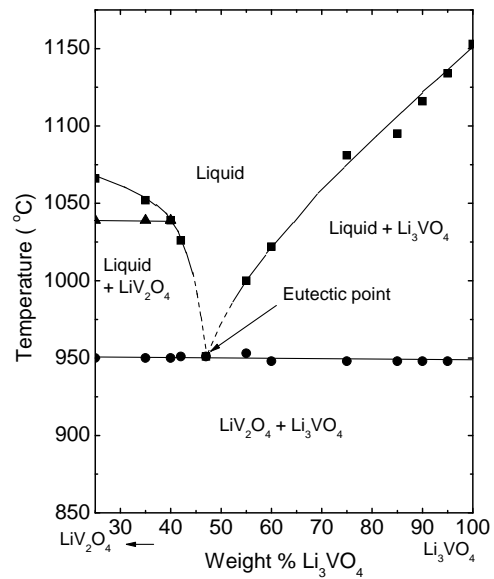}
\caption{Pseudobinary phase diagram of the LiV$_2$O$_4$-Li$_3$VO$_4$ system. The arrow below the horizontal axis points towards pure ${\rm LiV_2O_4}$, which is far to the left of the left-hand vertical axis.  Note that ${\rm LiV_2O_4}$ decomposes peritectically at 1040 $^\circ$C.}
\label{binphasediag}
\end{figure}

As a first step to find a self-flux for crystal growth of ${\rm LiV_2O_4}$, we melted a prereacted powder sample under inert atmosphere. The product was a mixture primarily of ${\rm V_2O_3}$ and ${\rm Li_3VO_4}$. Since the phase relations in the solid state at 700 $^\circ$C showed that ${\rm LiV_2O_4}$ is in equilibrium with both ${\rm V_2O_3}$ and ${\rm Li_3VO_4}$,\cite{Das} this result indicated that ${\rm Li_3VO_4}$ might be used as a flux to grow crystals of ${\rm LiV_2O_4}$. We therefore determined the ${\rm LiV_2O_4}$ -- ${\rm Li_3VO_4}$ pseudobinary phase diagram using a DTA under 1~atm He pressure, and the result is shown in Fig.~\ref{binphasediag}. We find that ${\rm LiV_2O_4}$ decomposes peritectically at 1040~$^\circ$C\@. This temperature is comparable to the maximum stability temperature of 1020~$^\circ$C for ${\rm LiV_2O_4}$ in vacuum found in Ref.~[\onlinecite{Matsushita}]. The eutectic temperature is about 950~$^\circ$C and the eutectic composition is approximately 53~wt\% ${\rm LiV_2O_4}$ and 47~wt\% ${\rm Li_3VO_4}$. We see from Fig.~\ref{binphasediag} that by cooling a liquid with a composition of 53 -- 58~wt\% of ${\rm LiV_2O_4}$ in ${\rm Li_3VO_4}$, crystals of ${\rm LiV_2O_4}$ should grow once the temperature reaches the liquidus temperature, until the growth temperature decreases to the eutectic temperature 950~$^\circ$C\@. Our flux ${\rm Li_3VO_4}$ has no other elements except Li, V and O, which eliminates the possibility of incorporating foreign elements in the ${\rm LiV_2O_4}$ crystals. Also ${\rm Li_3VO_4}$ did not evaporate at high temperatures $\sim$ 1000 -- 1100~$^\circ$C, nor did it react with or even wet platinum crucibles. The crystals could be separated from the flux by dissolving the flux in water (see below). All these data indicate that ${\rm Li_3VO_4}$ is an ideal flux for ${\rm LiV_2O_4}$ crystal growth.

\subsection{\label{growth}Crystal Growth}

Polycrystalline ${\rm LiV_2O_4}$ was prepared by the conventional solid state reaction of appropriate amounts of ${\rm Li_2CO_3}$, ${\rm V_2O_5}$ and ${\rm V_2O_3}$.\cite{Kondo1999} Powder samples of the flux ${\rm Li_3VO_4}$ were made by the solid state reaction of appropriate amounts of ${\rm V_2O_5}$ and ${\rm Li_2CO_3}$ at 525~$^\circ$C in air for $\sim$ 5 days. To grow the crystals we mixed powdered samples of ${\rm Li_3VO_4}$ and ${\rm LiV_2O_4}$ with the composition of 58~wt\% ${\rm LiV_2O_4}$ and 42~wt\% ${\rm Li_3VO_4}$. The mass of the ${\rm LiV_2O_4}$/${\rm Li_3VO_4}$ mixture was typically $\sim$ 5 -- 8 gm. The powder was placed in a platinum crucible which was then sealed under vacuum in a quartz tube. The quartz tube was then heated to 1038 -- 1060~$^\circ$C, was kept at that temperature for 12 -- 24 hours, and then cooled to 930~$^\circ$C at a slow rate. We obtained the largest (up to 2.5 mm on a side) crystals when the cooling rate was 1~$^\circ$C/hour. At higher cooling rates of 2~$^\circ$C/hour and 3~$^\circ$C/hour the crystal size became smaller (0.2 -- 0.5~mm on a side). From 930~$^\circ$C the sample was furnace-cooled to room temperature. The crystals of ${\rm LiV_2O_4}$ were extracted by dissolving the flux at 50 to 55~$^\circ$C in a solution of ${\rm LiVO_3}$ in deionized water or in pure deionized water in an ultrasonic bath. Finally the crystals were rinsed in acetone and dried. 

\begin{figure}
\includegraphics[width=2.5in]{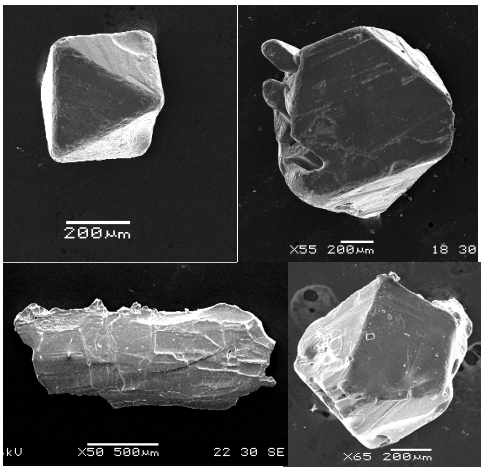}
\caption{Scanning electron microscope pictures of the three ${\rm LiV_2O_4}$ crystal morphologies obtained: octahedral (top left), irregular (top right and bottom right), and plate (bottom left).}
\label{pic}
\end{figure}

Three different kinds of crystal morphologies were obtained. One was octahedral shaped crystals with well-developed faces and size $\sim$ 1 mm on a side.  From Laue x-ray diffraction measurements, the flat faces of the octahedra were found to be [111] planes.  Another was irregular shaped: they were partly octahedral shaped with a few well-developed faces but also had irregular faces. Crystals with these two described morphologies were obtained together in the crystal growth runs. In one of our growth runs, along with the two morphologies,  some plate-shaped crystals were also obtained. These were $\sim$ 2 mm in length, $\sim$ 0.5 mm in width and about 0.1 mm in thickness. Figure \ref{pic} shows scanning electron microscope pictures of some of the crystals. X-ray diffraction measurements of powdered crystals showed single phase ${\rm LiV_2O_4}$. Some of the crystals were annealed at 700~$^\circ$C. To anneal, the crystals were wrapped in a platinum foil, embedded inside powder ${\rm LiV_2O_4}$ and then sealed in a quartz tube under vacuum. The presence of the powdered ${\rm LiV_2O_4}$ ensured that even trace amounts of oxygen in the tube would be taken up by the powder and the crystal would not become oxidized. 

\subsection{\label{structure}Chemical Analysis and Crystal Structure Determination}

Chemical analysis was carried out on a collection of $\sim$ 10 single crystals using inductively coupled plasma atomic emission spectroscopy (ICP-AES).\cite{Galbraith} The results gave the composition Li: (3.67 $\pm$ 0.37) wt\%; V: (54.6 $\pm$ 5.5) wt\%; O, by difference: (41.7 $\pm$ 5.9) wt\%. These values are consistent with the values calculated for ${\rm LiV_2O_4}$: Li, 4.0 wt\%; V, 59.0 wt\%; O, 37.0 wt\%.

A well-shaped octahedral crystal (0.25 $\times$ 0.25 $\times$ 0.21 mm$^3$) was selected for X-ray structure determination at $T$ = 293~K and $T$ = 193~K\@.  The initial cell constants were obtained from three series of $\omega$ scans at different starting angles. The final cell constants were calculated from a set of strong reflections from the actual data collection. The data were collected using the full sphere routine by collecting four sets of frames with 0.3$^\circ$ scans in $\omega$ with an exposure time 10 sec per frame with detector-to-crystal distance 3.5 cm\@. This data set was corrected for Lorentz and polarization effects. The absorption correction was based on fitting a function to the empirical transmission surface as sampled by multiple equivalent measurements.\cite{Blessing} X-ray structure determination and refinement were performed using SHELXTL software package.\cite{shelxtl}  
	
Cell parameters and the systematic absences in the diffraction data were consistent with the space group $F d\overline{3}m$ known for this compound. The Least-Squares refinement on $F^2$ converged to R1= 0.064 showing  significant extinction, therefore extinction correction was applied. The final results, presented in Tables~\ref{table1} and~\ref{table2}, are in a good agreement with earlier results on single crystals\cite{Matsushita} and powder diffraction.\cite{Chmaissem}

\begin{table*}
\caption{Crystal data and structure refinement of ${\rm LiV_2O_4}$. Here R1 = $\sum$$\mid$$\mid$$F$$_{\rm obs}$$\mid$~$-$~$\mid$$F$$_{\rm calc}$$\mid$$\mid$/$\sum$$\mid$$F$$_{\rm obs}$$\mid$ and wR2 = ($\sum$[ $w$($\mid$$F$$_{\rm obs}$$\mid$$^2$ $-$ $\mid$$F$$_{\rm calc}$$\mid$$^2$)$^2$]/$\sum$[ $w$($\mid$$F$$_{\rm obs}$$\mid$$^2$)$^2$])$^{1/2}$, where $F$$_{\rm obs}$ is the observed structure factor and $F$$_{\rm calc}$ is the calculated structure factor.}

\begin{ruledtabular}
\begin{tabular}{lll}
Temperature & 193(2) K & 293(2) K\\
\hline
Crystal system/Space group & Cubic, $F$$d$$\overline{3}$$m$ & Cubic, $F$$d$$\overline{3}$$m$\\
Unit cell parameter & $a$ = 8.2384(6) \AA & $a$ = 8.2427(7) \AA \\
Volume & 559.15(7) \AA$^3$ & 560.03(7) \AA$^3$ \\
Z & 8 & 8 \\
Density (Calculated) & 4.106 Mg/m$^3$ & 4.106 Mg/m$^3$\\
Absorption coefficient & 6.485 mm$^{-1}$ & 6.485 mm$^{-1}$ \\
F(000) & 648 & 648\\
Data / restraints / parameters & 80 / 0 / 8 & 80 / 0 / 8\\
Goodness-of-fit on $F$$^2$ & 1.392 & 1.401 \\
Final R indices [I $>$ 2$\sigma$(I)] & R1 = 0.0148 & R1 = 0.0180 \\
& wR2 = 0.0409 & wR2 = 0.0527\\ 
Extinction coefficient & 0.0205(15) & 0.0280(3)\\
\end{tabular}
\end{ruledtabular}
\label{table1}
\end{table*}

\begin{table}
\caption{Atomic coordinates ($\times$ 10$^4$) and equivalent isotropic displacement parameters ($\times$~10$^3$~\AA$^2$) for ${\rm LiV_2O_4}$ at 193 K\@. $U$(eq) is defined as one third of the trace of the orthogonalized $U$$_{\rm ij}$ tensor.}
\begin{ruledtabular}
\begin{tabular}{lllll}
 & x & y & z & $U$(eq)\\
\hline
V (1) & 5000 & 5000 & 5000 & 2(1)\\
O(1) & 2612(1) & 2612(1) & 2612(1) & 3(1)\\
Li (1) & 1250 & 1250 & 1250 & 2(2)\\
\end{tabular}
\end{ruledtabular}
\label{table2}
\end{table}

\section{\label{measurements}Physical Property Measurements}

\subsection{\label{magmeasurements}Magnetic Susceptibility}

\begin{figure}
\includegraphics[width=2.5in]{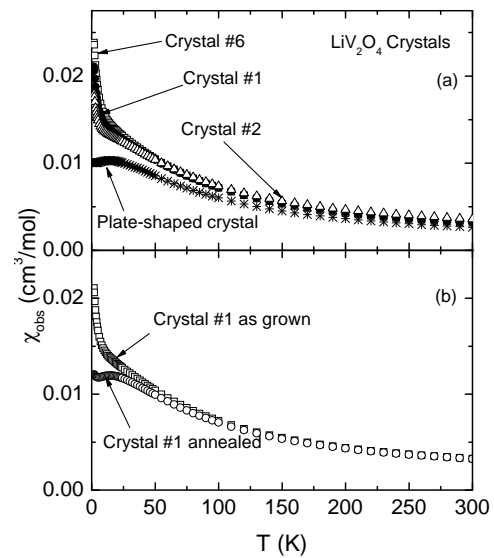}
\caption{(a) Observed magnetic susceptibility of an octahedral (crystal \#6), irregular (crystal \#1 and \#2) and a plate-shaped crystal of ${\rm LiV_2O_4}$. The sharp Curie-like upturn at low $T$ in the susceptibility of the octahedral and the irregular crystals show the presence of magnetic defects in the spinel structure of ${\rm LiV_2O_4}$. (b) Magnetic susceptibility of crystal \#1 (irregular shaped), as-grown and then annealed. The low $T$ sharp upturn for the as-grown crystal disappears after annealing at 700 $^\circ$C, showing the near-elimination of the magnetic defects by annealing.}
\label{susc}
\end{figure}

The magnetic susceptibilities of as-grown octahedral, irregular and plate-shaped crystals are shown in Fig.~\ref{susc}(a). The magnetic susceptibility of the octahedral and irregular crystals showed a sharp upturn at low temperatures which indicates that these as-grown crystals have magnetic defects in them, as also observed in some powder samples.\cite{Kondo1999} However the magnetic susceptibility of the plate-shaped crystal was strikingly different. The susceptibility of the plate showed only a tiny low-temperature upturn and therefore revealed the intrinsic susceptibility\cite{Kondo1999} of ${\rm LiV_2O_4}$. 

The magnetic defect concentration in a crystal was calculated by fitting the observed molar magnetization $M$${\rm _{obs}}$ isotherms at low temperatures\cite{Kondo1999} ($<$ 10 K, not shown) by the equation 
\begin{equation}
M{\rm _{obs}} = \chi H + n_{\rm defect} N{\rm _A}g{\rm _{defect}}\mu{\rm _B} S{\rm _{defect}} B_{S}(x)\, , 
\label{fiteq}
\end{equation}
where ${n_{\rm defect}}$ is the concentration of the magnetic defects, $N{\rm _A}$ Avogadro's number, $g{\rm _{defect}}$ the $g$-factor of the defect spins which was fixed to 2 (the detailed reasoning behind this is given in Ref.\ [\onlinecite{Kondo1999}]), $S{\rm _{defect}}$ the spin of the defects, $B _S(x)$ the Brillouin function, and $\chi$ the intrinsic susceptibility of ${\rm LiV_2O_4}$ spinel phase. The argument of the Brillouin function $B _S(x)$  is $x$~=~$g{\rm _{defect}}$$\mu{\rm _B}$$S{\rm _{defect}}$$H$/[k${\rm _B}$($T$$-$$\theta{\rm _{defect}}$)] where $\theta{\rm _{defect}}$ is the Weiss temperature associated with the magnetic defects. Using the above analysis we obtained $n$${\rm _{defect}}$ $\sim$ 0.25 -- 0.5~ mol\% for the as-grown octahedral and irregular crystals and $n$${\rm _{defect}}$ $\lesssim$ 0.01 mol\% for the plate-shaped crystals. 

In some of the octahedral/irregular crystals, annealing at 700~$^\circ$C led to the near-elimination of the magnetic defects. The magnetic susceptibilities of one of the irregular shaped crystals, as-grown and then annealed, are shown in Fig.~\ref{susc}(b). The low $T$ Curie-like upturn in the susceptibility for the as-grown crystal disappeared after annealing the crystal at 700~$^\circ$C for five days, with the susceptibility becoming almost $T$-independent at low $T$, revealing the near-elimination of the magnetic defects. For the as-grown crystal we found $n$${\rm _{defect}}$ = 0.38 mol\% and after annealing, the defect concentration $n$${\rm _{defect}}$ became 0.01 mol\%.

\subsection{\label{transport}Heat Capacity and Electrical Resistivity Measurements}

\begin{figure}
\includegraphics[width=2.5in]{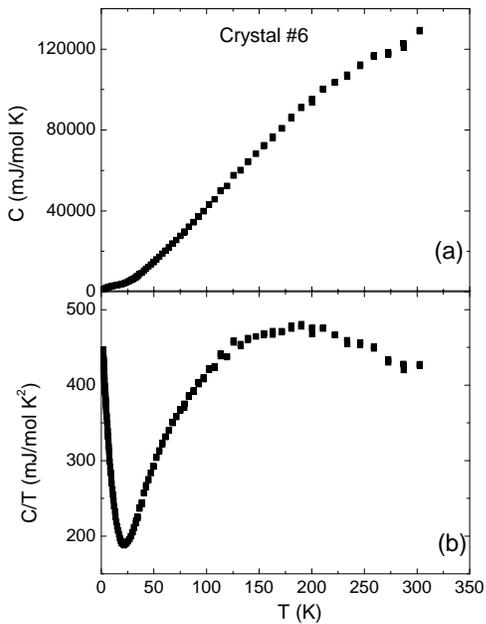}
\caption{(a) Specific heat $C$ versus temperature $T$ for an octahedral crystal of ${\rm LiV_2O_4}$ with magnetic defect concentration $n$${\rm _{defect}}$ = 0.5 mol\%. (b) The data in (a) plotted as $C/T$ versus $T$.}
\label{hc}
\end{figure}

\begin{figure}
\includegraphics[width=2.5in]{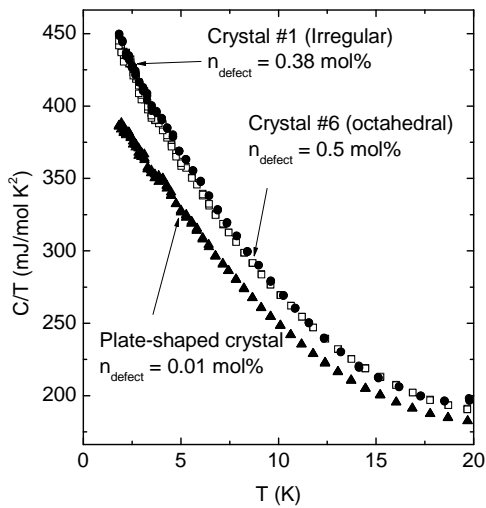}
\caption{$C/T$ versus $T$ for crystals with varying $n$${\rm _{defect}}$: Crystal \#1 (filled circles), crystal \#6 (open squares), and a plate-shaped crystal (filled triangles).}
\label{gamma}
\end{figure}

Figure \ref{hc}(a) shows the heat capacity $C$ and Fig.~\ref{hc}(b) shows the ratio $\gamma$ = $C/T$ of as-grown octahedral crystal \#6 [see also Fig.~\ref{susc}(a)]. Below 20 K, the $\gamma$ increases with decreasing $T$ and at the lowest temperature (1.8~K), it has a very high value of 450~mJ/mol~K$^2$, comparable to the values\cite{Johnston1999} of 420--430~mJ/mol~K$^2$ measured for powders.  Figure \ref{gamma} shows the low $T$ $\gamma$($T$) of an octahedral, an irregular and a plate-shaped crystal with different magnetic defect concentrations. The variations of $\gamma$ with $T$ for the octahedral and the irregular crystals are very similiar with the same value of $\gamma$ at the lowest temperature. However, $\gamma$ for the plate-shaped crystal is lower (380~mJ/mol~K$^2$) at 1.8~K\@. 

\begin{figure}
\includegraphics[width=2.5in]{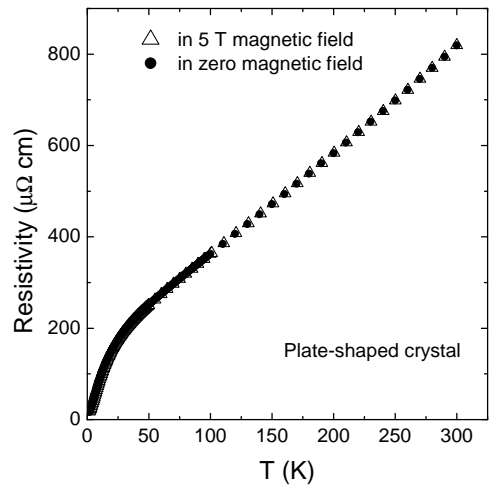}
\caption{Temperature $T$ variation of the resistivity of a plate-shaped crystal in zero field and in a 5 T magnetic field.}
\label{res}
\end{figure}

Figure \ref{res} shows the temperature variation of the four-probe resistivity of a plate-shaped crystal both in zero magnetic field and in 5 T magnetic field. The applied field of 5 T is seen to have little influence on the resistivity. The resistivity decreases with decreasing $T$ as expected for a metal. The residual resistivity ratio (RRR) for the plate-shaped crystal is 50, revealing its high crystal perfection. This value can be compared to the values of $\approx$~2, $\approx$~27 and $\approx$~12 for the crystals in Refs.~[\onlinecite{Rogers}], [\onlinecite{Takagi}] and [\onlinecite{Matsushita}], respectively. 

\section{\label{watertrmt}Water treatment of ${\rm \bf {LiV_2O_4}}$}

In view of the results in Ref.~[\onlinecite{Matsushita}], we also performed an experiment to see if our crystals of ${\rm LiV_2O_4}$ are sensitive to water exposure. We performed this experiment on both powdered samples and single crystals of ${\rm LiV_2O_4}$. We selected a sample of ${\rm LiV_2O_4}$ powder free of any magnetic defects. Then we put some of that powder into deionized water and some into a solution of ${\rm LiVO_3}$ in deionized water for two weeks. The X-ray diffraction patterns of the two treated samples remained unchanged from the original sample. The magnetic susceptibilities of the two treated samples along with that of the original sample are plotted in Fig.~\ref{watertrmt}(a). The susceptibilities of the three samples are nearly identical over the entire temperature range. With the single crystals, before dissolving the flux, a small crystal was broken off of the solified button of crystals embedded in the flux. The magnetic susceptibility of that small crystal was measured. Then it was put in water in an ultrasonic bath to dissolve the flux and after it was dried with acetone, it was left in water for 5 days. The magnetic susceptibility of that crystal before and after water treatment is shown in Fig.~\ref{watertrmt}(b). Our findings for both powder and single crystal ${\rm LiV_2O_4}$ contradict the results in Ref.~[\onlinecite{Matsushita}] where the susceptibility of their ${\rm LiV_2O_4}$ single crystals changes drastically after being exposed to air and moisture.

\begin{figure}
\includegraphics[width=2.5in]{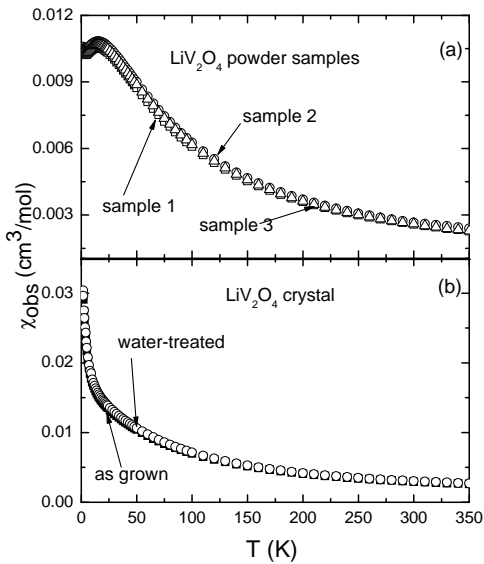}
\caption{(a) Magnetic susceptibility of three ${\rm LiV_2O_4}$ powder samples. Sample 1 (open squares) is the as-made ${\rm LiV_2O_4}$ sample while samples 2 (open circles) and 3 (open triangles) are water treated and ${\rm LiVO_3}$ solution treated, respectively. (b) Magnetic susceptibility of an as-grown (filled squares) and water treated (open circles) octahedral ${\rm LiV_2O_4}$ single crystal.}
\label{watertrmt}
\end{figure} 

\section{\label{summary}summary}

In this paper we have reported a new self-flux growth method to grow single crystals of ${\rm LiV_2O_4}$ using the flux ${\rm Li_3VO_4}$. The selection of ${\rm Li_3VO_4}$ as the flux led to the study of the ${\rm LiV_2O_4}$ -- ${\rm Li_3VO_4}$ pseudobinary phase diagram. ${\rm LiV_2O_4}$ was found to decompose peritectically at 1040 $^\circ$C. A eutectic was found with a eutectic temperature of 950~$^\circ$C and the eutectic composition being approximately 53~wt\% ${\rm LiV_2O_4}$ and 47~wt\% ${\rm Li_3VO_4}$. The crystals are of high quality, and as with powder ${\rm LiV_2O_4}$, are not reactive to air and moisture. The magnetic susceptibility of some of the crystals showed a Curie-like upturn at low $T$ showing the presence of magnetic defects within the spinel structure. The defects could be nearly eliminated in some of the crystals by annealing them at 700~$^\circ$C in vacuum. From the heat capacity measurements, a very large value of 450~mJ/mole~K$^2$ was obtained for $C$/$T$ for crystals having magnetic defects in them while a value of 380~mJ/mol~K$^2$ was obtained for crystals which were free of any magnetic defects. The electrical resistivity measurement on a magnetically pure crystal revealed the expected metallic behavior down to 1.8~K. 

In addition to the further study of heavy fermion behaviors in magnetically pure LiV$_2$O$_4$, the present method of crystal growth opens up new research areas associated with the physics of magnetic defects in LiV$_2$O$_4$.  From detailed high resolution electron diffraction and/or synchrotron x-ray structural studies one may be able to determine the nature of the crystal defects giving rise to the magnetic defects.  Important fundamental issues that can be addressed include whether the heavy Fermi liquid in magnetically pure LiV$_2$O$_4$ survives when magnetic defects are present and whether the crystal and magnetic defects drive a metal-insulator transition at some defect concentration.  These questions can initially be addressed in the milliKelvin temperature range using electrical resistivity, magnetic susceptibility, NMR, and electrical resistivity measurements.  A related question is whether a quantum critical point occurs versus magnetic defect concentration.  These are exciting topics for future research.

\begin{acknowledgments}
We thank A.~Kreyssig and S.~Nandi for x-ray Laue diffraction measurements of our octahedral crystals.  Work at the Ames Laboratory was supported by the Department of Energy-Basic Energy Sciences under Contract No.~DE-AC02-07CH11358.
\end{acknowledgments}

\end{document}